\documentclass[a4paper,fleqn]{cas-dc}
\usepackage{amsmath,amssymb,amsfonts}
\usepackage{algorithmic}
\usepackage{graphicx}
\usepackage{textcomp}
\usepackage{algorithm}
\usepackage{xcolor}
\usepackage{booktabs}
\usepackage[authoryear]{natbib}

\usepackage{subfiles}
\usepackage{tikz}
\usepackage[subpreambles=true]{standalone}
\usetikzlibrary{fit}
\usetikzlibrary{positioning}
\usetikzlibrary{quotes, arrows.meta}
\usepackage{deps/plotnueralnet/Box}
\usepackage{deps/plotnueralnet/RightBandedBox}
\setcitestyle{numbers}

\def\tsc#1{\csdef{#1}{\textsc{\lowercase{#1}}\xspace}}
\tsc{WGM}
\tsc{QE}
\tsc{EP}
\tsc{PMS}
\tsc{BEC}
\tsc{DE}

\begin{document}
\let\WriteBookmarks\relax
\def\floatpagepagefraction{1}
\def\textpagefraction{.001}

\shorttitle{Schwehr Z. and Achanta S.}
\shortauthors{Schwehr Z. and Achanta S.}

\title [mode = title]{Brain Tumor Segmentation Based on Deep Learning, Attention Mechanisms, and Energy-Based Uncertainty Prediction}                      
\tnotemark[1]
\tnotetext[1]{This research did not receive any specific grant from funding agencies in the public, commercial, or not-for-profit sectors.}

\author[1]{Zachary Schwehr}[orcid=0009-0005-4440-9881]
\cormark[1]
\ead{zschwehr1@gmail.com}
\credit{Conceptualization of this study, Methodology, Software, Editing}

\affiliation[1]{organization={Mills E. Godwin High School},
    addressline={2101 Pump Road}, 
    city={Henrico},
    postcode={23238}, 
    state={Virginia},
    country={United States}}

\author[1]{Sriman Achanta}[orcid=0009-0006-4686-2652]
\ead{achantass25@gmail.com}
\credit{Software, Editing}

\cortext[cor1]{Corresponding author}

\begin{abstract}
Brain tumors are one of the deadliest forms of cancer with a mortality rate of over 80\%. A quick and accurate diagnosis is crucial to increase the chance of survival. However, in medical analysis, manual annotation and segmentation of a brain tumor can be complicated. Multiple MRI modalities are typically analyzed as they provide unique information regarding the tumor regions. Although these MRI modalities are helpful for segmenting gliomas, they tend to increase overfitting and computation. This paper proposes a region of interest detection algorithm that is implemented during data preprocessing to locate salient features and remove extraneous MRI data. This decreases the input size, allowing for more aggressive data augmentations and deeper neural networks. Following the preprocessing of the MRI modalities, a fully convolutional autoencoder segments the different brain MRIs which employs channel-wise self-attention and attention gates. Subsequently, test time augmentations and an energy-based model were used for voxel-based uncertainty predictions. Experimentation was conducted on the BraTS 19, 20, and 21 benchmarks, and the proposed models achieved state-of-the-art segmentation performance with mean dice scores of 84.55, 88.52, and 90.82 on each dataset, respectively. Additionally, qualitative results were used to assess the segmentation models and uncertainty predictions. The code for this work is made available online at: \url{https://github.com/WeToTheMoon/BrainTumorSegmentation}.
\end{abstract}

\begin{highlights}
\item State-of-the-art brain tumor segmentation on BraTS benchmarks consisting of high and low-grade gliomas
\item Hard and soft attention improves performance and effectively counteracts class imbalance
\item Test-time augmentations and energy-based model models have shown to effectively quantify uncertainty in semantic segmentation tasks
\end{highlights}


\maketitle

\section{Introduction}
Brain tumors, foreign cell growths in the brain, are one of the deadliest forms of cancer responsible for killing over 250,000 people worldwide annually, with a mortality rate of over 80\% \cite{b1}. Gliomas, the most prevalent type of brain tumor, occur due to the carcinogenesis of glial cells in the brain. Gliomas have a median survival time of 15 months with roughly 5\% of patients surviving after 5 years \cite{b2}. Magnetic Resonance Imaging (MRI) is a medical imaging technique employed by physicians to diagnose and treat brain tumors through the use of various image modalities. However, the manual annotation and analysis of the different 3D MRI modalities is an arduous and time-consuming task that can only be performed by neuroradiologists. A proper brain tumor diagnosis takes neuroradiologists several hours with an error rate of nearly 20\% \cite{b3,b4}. Therefore, an automated method of segmenting brain tumors will heavily aid in the diagnosis and treatment of brain tumors. The predicted lesion segmentation provides neuroradiologists with key information regarding the size, location, and shape of the various tumor regions (enhancing tumor, peritumoral edema, and the necrotic and non-enhancing tumor core) which allows for a more efficient and effective treatment. 

\quad Recently, several models have been proposed in literature for medical segmentation. These proposed models can be separated into two major groups: machine learning and deep learning methods. Deep learning methods are a subsection of machine learning as they attempt to recreate the synapses in the brain to learn patterns and features in the data. Both these techniques take learned features and patterns from previous training data, however, deep learning achieves far greater results given larger training datasets. Common examples of machine learning techniques include random forest, support vector machine (SVM), and clustering.  These methods have yielded strong results in the field of medical imaging, e.g., classification, disease detection, and segmentation. However, when working with complex data where regions are vague and difficult to segment, these methods tend to perform poorly. Previous literature attempts at brain tumor segmentation are relatively inaccurate when segmenting the borders of the different regions. Additionally, they suffer from a class imbalance where the regions with a small area tend to have poor performance. Furthermore, previous algorithms are unable to effectively quantify the model’s uncertainty. The two primary types of uncertainty are aleatoric and epistemic. Aleatoric and epistemic uncertainty are caused by random noise and a lack of knowledge of the model, respectively \cite{b5}. As a result, physicians are unable to decipher between accurate and inaccurate predictions, which decreases the credibility of the model. This prevents an effective clinical implementation as the predictions are often ignored due to their fallacies.

\quad To account for the shortcomings mentioned above, this paper implements aggressive preprocessing techniques and a robust multiclass segmentation model to increase segmentation performance and reduce class imbalance. The preprocessing techniques function by extracting the region of interest from the MRI modalities, and cropping out extraneous information. This aggressive preprocessing allows the model to focus on the tumor regions. Additionally, the reduced input size decreases computational costs, allowing for a deeper and more robust model architecture. To quantify uncertainty, we employed test-time augmentations and an energy-based model for voxel-based uncertainty predictions. These methods create a visual confidence map that is simple for physicians to understand. Additionally, the voxel-based uncertainty predictions allow analysts to determine the uncertainty of a specific region in addition to the prediction itself.

\section{Related Work}
\textbf{Medical Image Segmentation:} Most medical segmentation models are neural networks, primarily convolutional neural networks (CNN). These neural networks include fully convolutional networks (FCN) \cite{b6} and autoencoders that are based on the U-Net \cite{b7}. Modified architectures of these models have been used for the semantic segmentation of both medical and non-medical images. Semantic segmentation assigns each pixel or voxel in the image with a given class or label. For medical imaging, the Attention U-Net was built upon the U-Net’s architecture by incorporating attention gates. These gates highlight important information while ignoring irrelevant information within the skip connections of the U-Net. The Attention U-Net performed well in segmenting the pancreas, highlighting the capabilities of soft attention mechanisms \cite{b8}. The UNet++ is another modified U-Net that was proposed by Z. Zhou et al. The UNet++ modifies the U-Net’s skip connections by implementing a denser convolution block that replaces the skip connection. The outputs from the convolution blocks are then concatenated with the previous decoded node and the corresponding encoder node. These modified skip connection pathways increased the model’s accuracy, with the downside of additional parameters and greater computational expenses \cite{b9}. In addition to the U-Net, the cascaded convolutional neural network has shown promising results in the field of medical imaging. H. Roth et al implemented a cascaded convolutional neural network to segment organs and the major vessels in the abdomen. The cascaded convolutional neural network consists of two stages. The first stage of the neural network outputs a mask of the human body. This mask is then used in the second stage to compute the final prediction. This method heavily improved the mean dice score from 68.5 to 82.2\% \cite{b10}. 

\quad \textbf{Brain Tumor Segmentation:} The task of brain tumor segmentation aims to accurately segment the brain into different brain tumor regions. These regions include normal brain tissue, peritumoral edema, enhancing tumor region, and the necrotic and non-enhancing tumor region. In this paper, we propose various methods for brain tumor segmentation on the BraTS 2019, 2020, and 2021 datasets; these datasets contain multi-modality brain MRIs composed of various slices. The task of brain tumor segmentation is particularly difficult due to its high dimensionality, as well as the high variation between different MRI scans: varying shape, size, and location. Additionally, the low quality and clarity of the brain MRIs lead to increased complexity. Previous literature has explored if the aggregation of the 2D slices yields greater accuracy than the 3D MRI. Avesta et al compared 2D slices, 2.5D (5 consecutive slices), and 3D brain MRIs for segmenting the different anatomical parts of the brain (3rd Ventricle, Thalamus, and Hippocampus). They discovered that the 3D brain MRIs far outperformed the 2D and 2.5D inputs. However, the 3D inputs required more memory for training
\cite{b11}.

\quad \textbf{Attention Mechanism:} Attention mechanisms have often been implemented in deep learning to improve the performance of the model. These models learn attention or the importance of the different inputs allowing the model to better understand the data. Attention was largely introduced in the transformer which was proposed in the field of natural language processing \cite{b45}. The transformer employs attention mechanisms, allowing the model to learn the relationship and importance of the different words leading to massive performance improvements. Furthermore, attention mechanisms have been employed in many other fields such as computer vision or geometric deep learning tasks. The two major types of attention mechanisms include soft and hard attention. In soft attention, data is assigned attention weights or coefficients, allowing the model to be fully differentiable such as in the transformer. Hard attention similarly maps the relationship between the data, however, the mechanism is not differentiable, preventing the model from learning using gradient descent \cite{b12}. 

\quad \textbf{Uncertainty Estimations.} The primary methods to quantify uncertainty for neural networks are simple neural deterministic, Bayesian, ensemble, and test-time augmentation. A simple neural deterministic model quantifies uncertainty through an external model or is predicted by the same neural network. The two most common probability density functions in neural networks include the softmax and sigmoid activation functions, which are commonly interpreted as uncertainty estimations. However, these functions are often overconfident and commonly yield inaccurate uncertainty estimations. As a result, various other models have been implemented such as the Dirichlet distribution. Another method for uncertainty estimation is the aggregation of various predictions. For example, Bayesian methods include altering the parameters of the predictive model and then combining the different predictions after changing the weights. Alternate examples include combining the predictions of an ensemble and aggregating the predictions of the different augmentations of the same image \cite{b13}. 

\section{Methods}

\subsection{Pre-Processing}
In this study, we implement aggressive data pre-processing techniques to remove irrelevant information and decrease overfitting, the number of false negatives, and computation. To yield more accurate predictions, we utilized 4 MRI modalities of the brain: T1, T1-Gd, T2, and FLAIR. The T1 and T1-Gd with gadolinium-enhancing contrast modalities highlight fat locations that represent the enhancing and non-enhancing regions. The T2 and Fluid Attenuated Inversion Recovery (FLAIR) modalities show the presence of fluids and water which is useful in detecting peritumoral edema. As a result, all four MRI modalities are needed to accurately segment the different tumor regions due to the dependence on the presence of vascularity and fluids within the brain. To make the data uniform and remove any anisotropic effects, we implemented Z-score normalization which subtracted each voxel by the mean and divided by the standard deviation, resulting in a zero mean and unit variance. The formula for Z-Score normalization can be defined as:

\begin{equation}
    \hat{x_{i}} = \frac{x_{i} - \mu}{\sigma}
\end{equation}

\noindent This step decreases overfitting and the occurrence of abnormal data which allows for faster convergence \cite{b14}.

\subsection{Region of Interest Detection Algorithm}

All brain MRIs contain information that is not pertinent to the diagnosis of the brain tumor. This includes the healthy brain tissue and the void regions of the MRI. Since this information is not important for the model, we implemented a hard attention algorithm to crop the MRIs along the x and y-axis. To locate these salient features and crop out extraneous MRI information, we use a binary segmentation model to locate the regions in the brain that are affected by the brain tumor, including the regions of peritumoral edema, enhancing tumoral regions, and non-enhancing tumoral regions. The four 3D MRI modalities were cropped to the dimensions of the binary mask plus a 12-voxel tolerance to account for potential errors in the binary model. Additionally, cropped MRIs with dimensions below the threshold of $48 \times 48 \times 128$ were additionally padded to such dimensions to allow for proper patch extraction.

\begin{figure*}
    \begin{center}
        \begin{tikzpicture}[scale=0.55, transform shape]
  \tikzstyle{connection}=[ultra thick,every node/.style={sloped,allow upside down},draw=\edgecolor,opacity=0.7]
  \tikzstyle{copyconnection}=[ultra thick,every node/.style={sloped,allow upside down},draw={rgb:blue,4;red,1;green,1;black,3},opacity=0.7]

  % First Conv Block
  \pic[shift={(0,0,0)}] at (0,0,0) {RightBandedBox={name=cr0,%
        fill=\BinaryConvColor,bandfill=\BinaryConvReluColor,
        height=48,width={2},depth=48}};
  \pic[shift={(0,0,0)}] at (cr0-east) {Box={name=p0,%
        fill=\BinaryPoolColor,opacity=0.5,height=40,width=1,depth=40}};

  % Second Conv Block
  \pic[shift={(1.5,0,0)}] at (p0-east) {RightBandedBox={name=cr1,%
       fill=\BinaryConvColor,bandfill=\BinaryConvReluColor,%
        height=40,width={2},depth=40}};
  \pic[shift={(0,0,0)}] at (cr1-east) {Box={name=p1,%
        fill=\BinaryPoolColor,opacity=0.5,height=32,width=1,depth=32}};
  
  % Third Conv Block
  \pic[shift={(1.5,0,0)}] at (p1-east) {RightBandedBox={name=cr2,%
        fill=\BinaryConvColor,bandfill=\BinaryConvReluColor,%
        height=32,width={3.5},depth=32}};
  \pic[shift={(0,0,0)}] at (cr2-east) {Box={name=p2,%
        fill=\BinaryPoolColor,opacity=0.5,height=25,width=1,depth=25}};

  % Fourth Conv Block
  \pic[shift={(1.25,0,0)}] at (p2-east) {RightBandedBox={name=cr3,%
        fill=\BinaryConvColor,bandfill=\BinaryConvReluColor,%
        height=25,width={4.5},depth=25}};
  \pic[shift={(0,0,0)}] at (cr3-east) {Box={name=p3,%
        fill=\BinaryPoolColor,opacity=0.5,height=16,width=1,depth=16}};

  % Blacks my favorite color
  \pic[shift={(1.25,0,0)}] at (p3-east) {RightBandedBox={name=cr4, fill=\BinaryConvColor,bandfill=\BinaryConvReluColor,%
        height=18,width={5},depth=18}};

  % Transpose Shit

  \pic[shift={(1.5, 0,0)}] at (cr4-east) {Box={name=transpose1,%
        fill=\BinaryConvTransposeColor,%
        height=25,width=2,depth=25}};
  \pic[shift={(0,0,0)}] at (transpose1-east) {Box={name=concat1,%
        fill=\BinaryConcatenateColor,opacity=0.5,height=25,width=4,depth=25}};
  \pic[shift={(0, 0,0)}] at (concat1-east) {RightBandedBox={name=ucr3,%
        fill=\BinaryConvColor,bandfill=\BinaryConvReluColor,%
        height=25,width=2,depth=25}};

  \pic[shift={(1.5, 0,0)}] at (ucr3-east) {Box={name=transpose2,%
        fill=\BinaryConvTransposeColor,%
        height=32,width=2,depth=32}};
  \pic[shift={(0,0,0)}] at (transpose2-east) {Box={name=concat2,%
        fill=\BinaryConcatenateColor,opacity=0.5,height=32,width=4,depth=32}};
  \pic[shift={(0, 0,0)}] at (concat2-east) {RightBandedBox={name=ucr2,%
        fill=\BinaryConvColor,bandfill=\BinaryConvReluColor,%
        height=32,width=2,depth=32}};

  \pic[shift={(1.5, 0,0)}] at (ucr2-east) {Box={name=transpose3,%
        fill=\BinaryConvTransposeColor,%
        height=40,width=2,depth=40}};
  \pic[shift={(0,0,0)}] at (transpose3-east) {Box={name=concat3,%
        fill=\BinaryConcatenateColor,opacity=0.5,height=40,width=4,depth=40}};
  \pic[shift={(0, 0,0)}] at (concat3-east) {RightBandedBox={name=ucr1,%
        fill=\BinaryConvColor,bandfill=\BinaryConvReluColor,%
        height=40,width=2,depth=40}};

  \pic[shift={(1.5, 0,0)}] at (ucr1-east) {Box={name=transpose4,%
        fill=\BinaryConvTransposeColor,%
        height=48,width=2,depth=48}};
  \pic[shift={(0,0,0)}] at (transpose4-east) {Box={name=concat4,%
        fill=\BinaryConcatenateColor,opacity=0.5,height=48,width=2,depth=48}};
  \pic[shift={(0, 0,0)}] at (concat4-east) {RightBandedBox={name=ucr0,%
        fill=\BinaryConvColor,bandfill=\BinaryConvReluColor,%
        height=48,width=2,depth=48}};

  \pic[shift={(0.75,0,0)}] at (ucr0-east) {Box={name=out,%
        fill=\BinarySoftmaxColor,height=49,width=1,depth=49}};

  \draw [connection]  (p1-east) -- node {\midarrow} (cr2-west);
  \draw [connection]  (p2-east) -- node {\midarrow} (cr3-west);
  \draw [connection]  (p3-east) -- node {\midarrow} (cr4-west);
  \draw [connection]  (cr4-east) -- node {\midarrow} (transpose1-west);
  \draw [connection]  (ucr3-east) -- node {\midarrow} (transpose2-west);
  \draw [connection]  (ucr2-east) -- node {\midarrow} (transpose3-west);
  \draw [connection]  (ucr1-east) -- node {\midarrow} (out-west);

  \path (cr3-southeast) -- (cr3-northeast) coordinate[pos=1.25] (cr3-top) ;
  \path (cr2-southeast) -- (cr2-northeast) coordinate[pos=1.25] (cr2-top) ;
  \path (cr1-southeast) -- (cr1-northeast) coordinate[pos=1.25] (cr1-top) ;

  \path (concat3-south)  -- (concat3-north)  coordinate[pos=1.25] (concat1-top) ;
  \path (concat2-south)  -- (concat2-north)  coordinate[pos=1.25] (concat2-top)  ;
  \path (concat1-south)  -- (concat1-north)  coordinate[pos=1.25] (concat3-top)  ;

  \draw [copyconnection]  (cr3-northeast)
  -- node {\copymidarrow}(cr3-top)
  -- node {\copymidarrow}(concat3-top)
  -- node {\copymidarrow} (concat1-north);

  \draw [copyconnection]  (cr2-northeast)
  -- node {\copymidarrow}(cr2-top)
  -- node {\copymidarrow}(concat2-top)
  -- node {\copymidarrow} (concat2-north);

  \draw [copyconnection]  (cr1-northeast)
  -- node {\copymidarrow}(cr1-top)
  -- node {\copymidarrow}(concat1-top)
  -- node {\copymidarrow} (concat3-north);
  
\tikzstyle{layerkey} = [
    rectangle, draw,
    minimum width=0.5cm, minimum height=0.5cm
]

  \node [layerkey, fill=\BinaryConvColor, shift={(2.65, 2.5)}] (r1) at (out-east) {};
  \node [layerkey, fill=\BinaryPoolColor, below=of r1] (r2) {};
  \node [layerkey, fill=\BinaryConvTransposeColor, below=of r2] (r3) {};
  \node [layerkey, fill=\BinaryConcatenateColor, below=of r3] (r4) {};
  \node [layerkey, fill=\BinarySoftmaxColor, below=of r4] (r5) {};

\node [right=0.35cm of r1, anchor=west] {ConvBlock1};
\node [right=0.35cm of r2, anchor=west] {ConvStrides};
\node [right=0.35cm of r3, anchor=west] {Transpose};
\node [right=0.35cm of r4, anchor=west] {Concatenate};
\node [right=0.35cm of r5, anchor=west] {Sigmoid};
\node[] at (-1.6,-7.1) {$128 \times 128 \times 128 \times 40$};
\node[] at (0.8,-6) {$64 \times 64 \times 64 \times 40$};
\node[] at (3.3,-4.9) {$32 \times 32 \times 32 \times 80$};
\node[] at (5.9,-3.9) {$16 \times 16 \times 16 \times 160$};
\node[] at (8.7,-2.9) {$8 \times 8 \times 8 \times 200$};
\node[] at (10.9,-3.9) {$16 \times 16 \times 16 \times 160$};
\node[] at (13.8,-4.9) {$32 \times 32 \times 32 \times 80$};
\node[] at (16.6,-6) {$64 \times 64 \times 64 \times 40$};
\node[] at (19.2,-7.1) {$128 \times 128 \times 128 \times 40$};
\end{tikzpicture}
    \end{center}
    \caption{\textbf{Fig. 1.} The architecture of the binary brain tumor segmentation model. This model is a U-Net-like structure with instance normalization, strided convolutions, and the ELU activation function. The strided convolutions and transposed convolutional layers have a scale of 2: decreasing and increasing the feature maps by a scale of 2, respectively. The input dimensions are $T \times 128 \times 128 \times 128 \times 4$ where $T$ is the batch size and the dimensions $128 \times 128 \times 128$ represent the size of the 3D brain MRI. There are four channels representing the MRI modalities, T1, T1-Gd, T2, and Flair. The output is $T \times 128 \times 128 \times 128$ representing the $T$ batches of the binarily segmented brain tumors.}
    \label{Fig:Model}
\end{figure*}
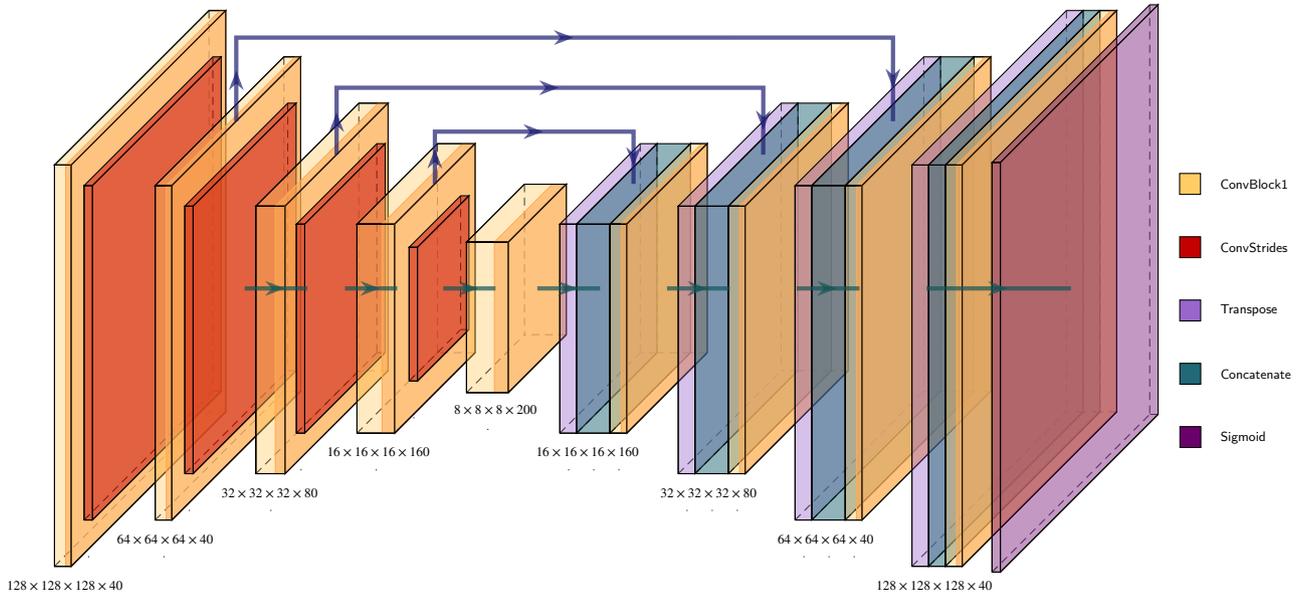

\subsection{Binary Segmentation Model}

The binary segmentation model predicts the binary mask of the brain tumor. The input dimensions for this model are $128 \times 128 \times 128 \times 4$ for the four 3D brain MRI modalities. All four MRI modalities are used as they are needed to properly segment the regions encompassing the brain tumor. Although peritumoral edema usually occurs on the extremities of the brain tumor, there are instances where the enhancing tumor regions are on the border of the brain tumor, requiring the use of all four MRI modalities. The output of the binary model is a binary segmentation mask with the same spatial dimensions as the input. This binary mask is used for cropping the brain tumor in the region of interest detection algorithm. The binary model is similar to the 3D U-Net, however, it employs a different number of feature channels and uses instance normalization and strided convolutions instead of max pooling. Max pooling takes the largest value within a certain kernel leading to data loss. As a result, we employ strided convolutions which decrease the size of the feature map while extracting information through convolutional layers. Additionally, we implement instance normalization between each convolutional layer to reduce overfitting and allow for faster convergence. The main difference between instance normalization and batch normalization is that instance normalization normalizes over a single image instead of the entire batch. Doing so prevents potential instance-specific shifts and decreases training noise caused by a limited batch size. The formula for instance normalization can be seen as:

\begin{equation}
\begin{aligned}
& y_{tijkc} = \frac{x_{tijkc}-\mu_{tc}}{\sqrt{\sigma^{2}_{tc}+\epsilon}} \\
& \mu_{tc} = \frac{1}{HWD}\sum_{i}^{H}\sum_{j}^{W}\sum_{k}^{D}x_{tijkc} \\ 
& \sigma^{2}_{tc} = \frac{1}{HWD}\sum_{i}^{H}\sum_{j}^{W}\sum_{k}^{D}(x_{tijkc}-\mu_{tc})
\end{aligned}
\end{equation}

\noindent where $x\in \mathbb{R}^{T\times H\times W\times D\times C}$ with a batch size of $T$ images, $H$, $W$, $D$, represent the height, width, and depth, respectively, and $\epsilon$ is a smoothing factor to prevent division by zero \cite{b15}.

\begin{figure}
    \begin{center}
        \begin{tikzpicture}[
    block/.style={
        rectangle, draw, minimum width=2cm, minimum height=0.75cm,
        rounded corners, thick
    },
    scale=0.9,
    transform shape
]
\definecolor{ConvColor}{RGB}{218,227,242}
\definecolor{ELUColor}{RGB}{250,226,191}
\definecolor{InstanceNormColor}{RGB}{205,202,240}

\node[block, rotate=90, fill=ConvColor] (conv1) at (0, 0) {3x3x3 Conv};
\node[block, rotate=90, fill=ELUColor, shift={(0,-1.5cm)}] (elu1) {ELU};
\node[block, rotate=90, fill=InstanceNormColor, shift={(0,-3cm)}] (in1) {I-Norm};
\node[block, rotate=90, fill=ConvColor, shift={(0,-4.5cm)}] (conv2) {3x3x3 Conv};
\node[block, rotate=90, fill=ELUColor, shift={(0,-6cm)}] (elu2) {ELU};
\node[block, rotate=90, fill=InstanceNormColor, shift={(0,-7.5cm)}] (in2) {I-Norm};

% \begin{scope}
%     \node [draw, dashed, rounded corners, inner sep=10pt, fit=(conv1) (elu1) (in1) (conv2) (elu2) (in2)] {};
% \end{scope}

\draw[-stealth, thick, shorten >=1pt] (conv1) -- (elu1);
\draw[-stealth, thick, shorten >=1pt] (elu1) -- (in1);
\draw[-stealth, thick, shorten >=1pt] (in1) -- (conv2);
\draw[-stealth, thick, shorten >=1pt] (conv2) -- (elu2);
\draw[-stealth, thick, shorten >=1pt] (elu2) -- (in2);
\end{tikzpicture}
    \end{center}
    \caption{\textbf{Fig. 2.} ConvBlock1 employs two convolutional layers with a kernel size of ($3 \times 3 \times 3$), instance normalization, and ELU.}
    \label{Fig:Model}
\end{figure}
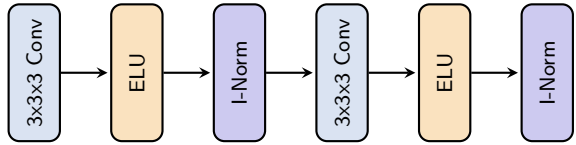
 
\quad The binary segmentation model is a fully convolutional autoencoder, including 3 regions: encoder, bridge, and decoder \cite{b7}. The structure of the multiclass model can be seen in Figure 1. The encoder and decoder portions each contain 4 blocks. Each encoder block contains 3 convolutional layers with instance normalization between each convolution. All of the  convolutional layers, except the transposed convolution, have a kernel size of $(3\times3\times3)$ with padding and used ELU as follows:

\begin{equation}
\begin{aligned}
f(x)=\begin{cases}
x, & x\ge 0 \\
\alpha (e^{x}-1), & x < 0 \\
\end{cases} \\
\end{aligned}
\end{equation}

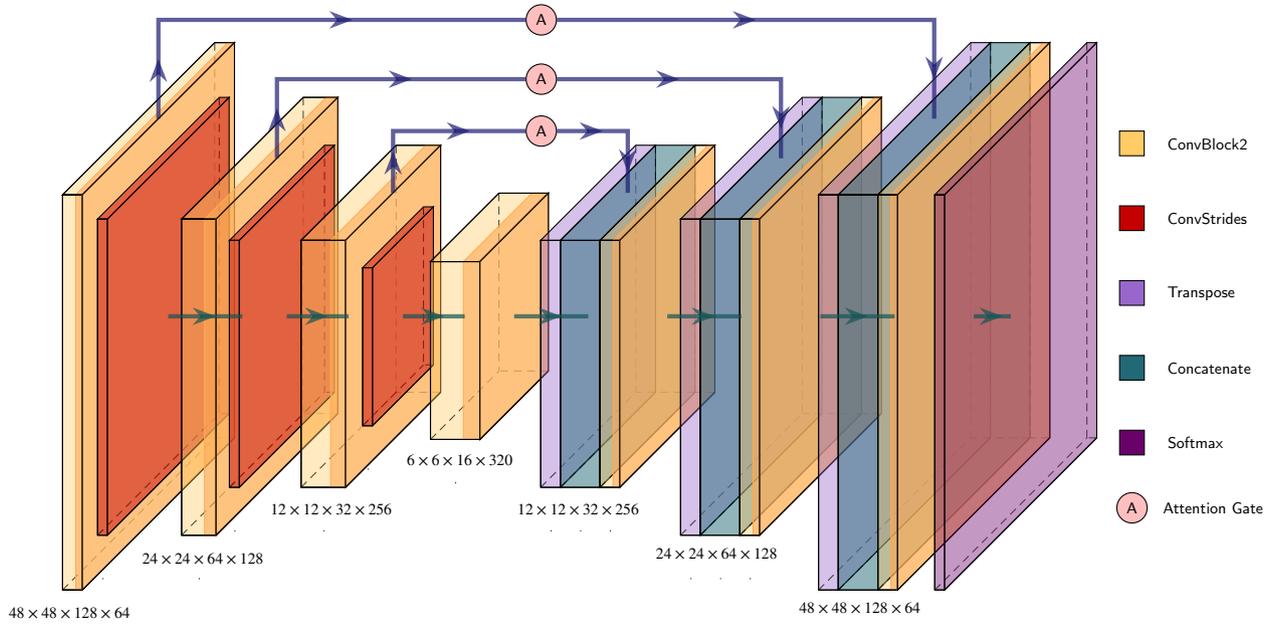
\begin{figure*}
    \begin{center}
        \begin{tikzpicture}[scale=0.65,transform shape]
  \tikzstyle{connection}=[ultra thick,every node/.style={sloped,allow upside down},draw=\edgecolor,opacity=0.7]
  \tikzstyle{copyconnection}=[ultra thick,every node/.style={sloped,allow upside down},draw={rgb:blue,4;red,1;green,1;black,3},opacity=0.7]

  % First Conv Block
  \pic[shift={(0,0,0)}] at (0,0,0) {RightBandedBox={name=cr1,%
        fill=\MulticlassConvColor,bandfill=\MulticlassConvReluColor,%
        height=40,width={2},depth=40}};
  \pic[shift={(0,0,0)}] at (cr1-east) {Box={name=p1,%
    fill=\MulticlassPoolColor,opacity=0.5,height=32,width=1,depth=32}};

  % Second Conv Block
  \pic[shift={(1.5,0,0)}] at (p1-east) {RightBandedBox={name=cr2,%
        fill=\MulticlassConvColor,bandfill=\MulticlassConvReluColor,%
        height=32,width={3.5},depth=32}};
  \pic[shift={(0,0,0)}] at (cr2-east) {Box={name=p2,%
    fill=\MulticlassPoolColor,opacity=0.5,height=25,width=1,depth=25}};

  % Third Conv Block
  \pic[shift={(1.25,0,0)}] at (p2-east) {RightBandedBox={name=cr3,%
        fill=\MulticlassConvColor,bandfill=\MulticlassConvReluColor,%
        height=25,width={4.5},depth=25}};
  \pic[shift={(0,0,0)}] at (cr3-east) {Box={name=p3,%
    fill=\MulticlassPoolColor,opacity=0.5,height=16,width=1,depth=16}};

  % Blacks my favorite color
  \pic[shift={(1.25,0,0)}] at (p3-east) {RightBandedBox={name=cr4,%
        fill=\MulticlassConvColor,bandfill=\MulticlassConvReluColor,%
        height=18,width={5},depth=18}};

  % Transpose Shit

  \pic[shift={(1.5, 0,0)}] at (cr4-east) {Box={name=transpose1,%
        fill=\MulticlassConvTransposeColor,%
        height=25,width=2,depth=25}};
  \pic[shift={(0,0,0)}] at (transpose1-east) {Box={name=concat1,%
fill=\MulticlassConcatenateColor,opacity=0.5,height=25,width=4,depth=25}};
  \pic[shift={(0, 0,0)}] at (concat1-east) {RightBandedBox={name=ucr3,%
        fill=\MulticlassConvColor,bandfill=\MulticlassConvReluColor,%
        height=25,width=2,depth=25}};

  \pic[shift={(1.5, 0,0)}] at (ucr3-east) {Box={name=transpose2,%
        fill=\MulticlassConvTransposeColor,%
        height=32,width=2,depth=32}};
  \pic[shift={(0,0,0)}] at (transpose2-east) {Box={name=concat2,%
fill=\MulticlassConcatenateColor,opacity=0.5,height=32,width=4,depth=32}};
  \pic[shift={(0, 0,0)}] at (concat2-east) {RightBandedBox={name=ucr2,%
        fill=\MulticlassConvColor,bandfill=\MulticlassConvReluColor,%
        height=32,width=2,depth=32}};

  \pic[shift={(1.5, 0,0)}] at (ucr2-east) {Box={name=transpose3,%
        fill=\MulticlassConvTransposeColor,%
        height=40,width=2,depth=40}};
  \pic[shift={(0,0,0)}] at (transpose3-east) {Box={name=concat3,%
        fill=\MulticlassConcatenateColor,opacity=0.5,height=40,width=4,depth=40}};
  \pic[shift={(0, 0,0)}] at (concat3-east) {RightBandedBox={name=ucr1,%
        fill=\MulticlassConvColor,bandfill=\MulticlassConvReluColor,%
        height=40,width=2,depth=40}};

  \pic[shift={(0.75,0,0)}] at (ucr1-east) {Box={name=out,%
        fill=\MulticlassSoftmaxColor,height=40,width=1,depth=40}};

  \draw [connection]  (p1-east) -- node {\midarrows} (cr2-west);
  \draw [connection]  (p2-east) -- node {\midarrows} (cr3-west);
  \draw [connection]  (p3-east) -- node {\midarrows} (cr4-west);
  \draw [connection]  (cr4-east) -- node {\midarrows} (transpose1-west);
  \draw [connection]  (ucr3-east) -- node {\midarrows} (transpose2-west);
  \draw [connection]  (ucr2-east) -- node {\midarrows} (transpose3-west);
  \draw [connection]  (ucr1-east) -- node {\midarrows} (out-west);

  \path (cr3-southeast) -- (cr3-northeast) coordinate[pos=1.25] (cr3-top) ;
  \path (cr2-southeast) -- (cr2-northeast) coordinate[pos=1.25] (cr2-top) ;
  \path (cr1-southeast) -- (cr1-northeast) coordinate[pos=1.25] (cr1-top) ;

  \path (concat3-south)  -- (concat3-north)  coordinate[pos=1.25] (concat1-top) ;
  \path (concat2-south)  -- (concat2-north)  coordinate[pos=1.25] (concat2-top)  ;
  \path (concat1-south)  -- (concat1-north)  coordinate[pos=1.25] (concat3-top)  ;

  % Attenion Gates
  \node[circle, draw, fill=pink, shift={(3, 0, 0)}] at (cr3-top) (attention1) {A};
  \node[circle, draw, fill=pink, shift={(5.35, 0, 0)}] at (cr2-top) (attention2) {A};
  \node[circle, draw, fill=pink, shift={(7.75, 0, 0)}] at (cr1-top) (attention3) {A};

  \draw [copyconnection]  (cr3-northeast)
  -- node {\copymidarrows}(cr3-top)
  -- node {\copymidarrows}(attention1)
  -- node {\copymidarrows}(concat3-top)
  -- node {\copymidarrows} (concat1-north);

  \draw [copyconnection]  (cr2-northeast)
  -- node {\copymidarrows}(cr2-top)
  -- node {\copymidarrows}(attention2)
  -- node {\copymidarrows}(concat2-top)
  -- node {\copymidarrows} (concat2-north);

  \draw [copyconnection]  (cr1-northeast)
  -- node {\copymidarrows}(cr1-top)
  -- node {\copymidarrows}(attention3)
  -- node {\copymidarrows}(concat1-top)
  -- node {\copymidarrows} (concat3-north);

\tikzstyle{layerkey} = [
    rectangle, draw,
    minimum width=0.5cm, minimum height=0.5cm
]

  \node [layerkey, fill=\MulticlassConvColor, shift={(2.25, 3.5)}] (r1) at (out-east) {};
  \node [layerkey, fill=\MulticlassPoolColor, below=of r1] (r2) {};
  \node [layerkey, fill=\MulticlassConvTransposeColor, below=of r2] (r3) {};
  \node [layerkey, fill=\MulticlassConcatenateColor, below=of r3] (r4) {};
  \node [layerkey, fill=\MulticlassSoftmaxColor, below=of r4] (r5) {};    
  \node[circle, draw, fill=pink, shift={(0, 0.25)}, below=of r5] (r6) {A};

\node [right=0.35cm of r1, anchor=west] {ConvBlock2};
\node [right=0.35cm of r2, anchor=west] {ConvStrides};
\node [right=0.35cm of r3, anchor=west] {Transpose};
\node [right=0.35cm of r4, anchor=west] {Concatenate};
\node [right=0.35cm of r5, anchor=west] {Softmax};
\node [right=0.2cm of r6, anchor=west] {Attention Gate};
\node[] at (-1.4,-6) {$48 \times 48 \times 128 \times 64$};
\node[] at (1.3,-4.9) {$24 \times 24 \times 64 \times 128$};
\node[] at (3.9,-3.9) {$12 \times 12 \times 32 \times 256$};
\node[] at (6.5,-2.9) {$6 \times 6 \times 16 \times 320$};
\node[] at (8.9,-3.9) {$12 \times 12 \times 32 \times 256$};
\node[] at (11.7,-4.8) {$24 \times 24 \times 64 \times 128$};
\node[] at (14.6,-5.9) {$48 \times 48 \times 128 \times 64$};
\end{tikzpicture}
    \end{center}
    \begin{center}
        \caption{\textbf{Fig 3.} The architecture of the multiclass segmentation model. It is similar to the binary segmentation model, however, it employs soft attention mechanisms, more filter channels, fewer encoding, and decoding blocks, and ReLU instead of ELU. The input dimensions are $T \times 48 \times 48 \times 128 \times 4$. The output is $T \times 48 \times 48 \times 128 \times 4$ representing the $T$ batches of the segmented brain tumors with the four channels representing the four classes: normal brain tissue, peritumoral edema, enhancing tumor region, and non-enhancing and necrotic tumor region.}
    \end{center}

    \label{Fig:Model}
\end{figure*}

\begin{figure*}
    \begin{center}
        \begin{tikzpicture}[
    block/.style={
        rectangle, draw, minimum width=2.5cm, minimum height=0.75cm,
        rounded corners, thick
    }
]

\definecolor{ConvColor}{RGB}{218,227,242}
\definecolor{ReLUColor}{RGB}{255, 165, 0}
\definecolor{InstanceNormColor}{RGB}{205,202,240}
\definecolor{OtherThingy}{RGB}{251,236,195}
\definecolor{GapColor}{RGB}{197,224,180}

\node[block, rotate=90, fill=ConvColor] (conv1) at (0, 0) {3x3x3 Conv};
\node[block, rotate=90, fill=ReLUColor, shift={(0,-1.5cm)}] (relu1) {ReLU};
\node[block, rotate=90, fill=InstanceNormColor, shift={(0,-3cm)}] (in1) {I-Norm};
\node[block, rotate=90, fill=ConvColor, shift={(0,-4.5cm)}] (conv2) {3x3x3 Conv};
\node[block, rotate=90, fill=ReLUColor, shift={(0,-6cm)}] (relu2) {ReLU};
\node[block, rotate=90, fill=InstanceNormColor, shift={(0,-7.5cm)}] (in2) {I-Norm};
\node[block, rotate=90, shift={(0,-9cm)}, fill=GapColor] (gap1) {GAP};
\node[block, rotate=90, fill=OtherThingy, shift={(0,-10.5cm)}] (wc1) {${W_{c}}$};
\node[block, shift={(12.25cm, 0)}, minimum width=1.25cm, minimum height=1.25cm] (sig1) {$\mathlarger{\mathlarger{\mathlarger{\mathlarger{\mathlarger{\sigma}}}}}$};
\node[draw, circle, thick, shift={(13.75cm, 0)}] (ew1) {$\times$};
\node[block, rotate=90, fill=InstanceNormColor, shift={(0,-15cm)}] (in3) {I-Norm};

\foreach \i/\j in {conv1/relu1, relu1/in1, in1/conv2, conv2/relu2, relu2/in2, in2/gap1, gap1/wc1, wc1/sig1, sig1/ew1, ew1/in3}
  \draw[-stealth, thick, shorten >=1pt] (\i) -- (\j);
  
\draw[-stealth, thick, shorten >=1pt]  (conv2) -- node {} ++(0,-2cm) -| (ew1) node[pos=0.25] {} node[pos=0.75] {};

\node [above=0cm of sig1, anchor=south] {Sigmoid};

\coordinate (f1) at (6, -2);

% \begin{scope}
%     \node [draw, dashed, rounded corners, inner sep=10pt, fit=(conv1) (relu1) (in1) (conv2) (relu2) (in2) (gap1) (wc1) (sig1) (ew1) (in3) (f1)] {};
% \end{scope}

\end{tikzpicture}
    \end{center}
    
        \caption{\textbf{Fig 4.} ConvBlock2 employs two convolutional layers with a kernel size of ($3 \times 3 \times 3$), instance normalization, ReLU, and a channel-based attention algorithm. The channel-based attention algorithm computes the attention coefficient, $\alpha$, which maps the importance of each of the channels where GAP takes the global average of each channel.}

    \label{Fig:Model}
\end{figure*}
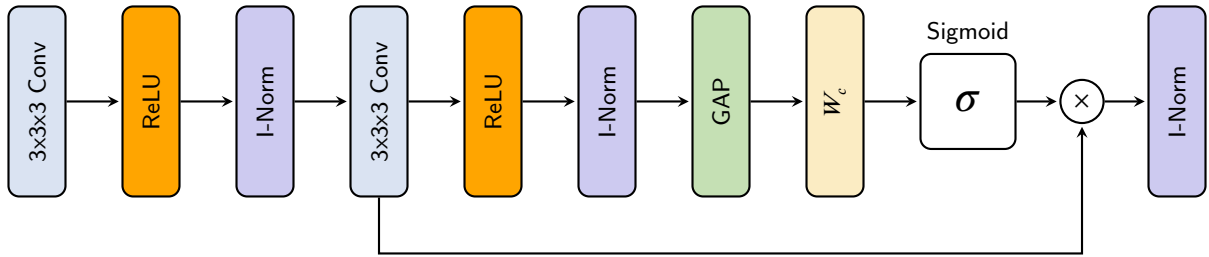

The third convolutional layer in each encoding block has a stride of $(2\times2\times2)$ to decrease the resolution. The bridge contains two convolutional layers with instance normalization. The decoding blocks consist of 3 convolutions with instance normalization, however, the first convolution transposes the data with a kernel size of $(2 \times 2 \times 2)$. Following the transposing convolutional layer, skip connections were implemented by concatenating the data. After the decoding region, a final convolutional layer with a kernel size of $(1\times1\times1)$ classifies the binary regions using the sigmoid activation function as follows:
\begin{equation}
\frac{1}{1+e^{-x}}
\end{equation}

\quad The model was trained with a learning rate of 0.0003 using the Adaptive Moment Estimation (Adam) optimizer and dice loss as in:
\begin{equation}
    \mathcal{L}_{DL} = DL(\hat{y}, y) = 1-\frac{\hat{y} \times y + \epsilon}{\hat{y}+y+\epsilon}
\end{equation}
where $\hat{y}$ and $y$ are the predicted and true masks, respectively. Additionally, a smooth factor, $\epsilon$, is added to prevent division by zero \cite{b17,b18}. The model was trained for 300 epochs with a batch size of 2.

Due to the limited training data, we employed data augmentations such as elastic deformation, random rotation, and random brightness during training. Elastic deformation moved each voxel to a new location and then used spline interpolation with an order of one to obtain the integer coordinates as follows: 
\begin{equation}
\begin{aligned}
& \Delta _{x} = G(\sigma)\times \text{Rand}(H, W, D) \\
& \Delta _{y} = G(\sigma)\times \text{Rand}(H, W, D) \\ 
& \Delta _{z} = G(\sigma)\times \text{Rand}(H, W, D) \\
& \hat{I}(i+\Delta _{x}(i,j,k),j+\Delta _{y}(i,j,k),k+\Delta _{z}(i,j,k)) \\
& = I(i,j,k) \\
\end{aligned}
\end{equation}

\noindent where $G(\sigma)$ is the Gaussian filter $(\sigma)$, and $H\times W\times D$ are the dimensions for each MRI \cite{b19}. During the training of the binary model, we set $\sigma = 2$. 

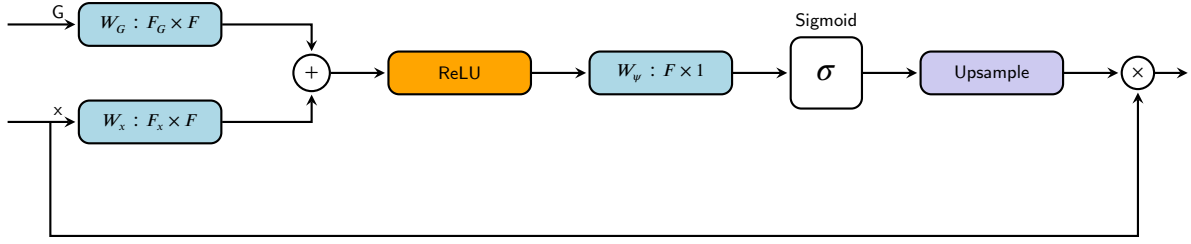
\begin{figure*}
    \begin{center}
        \begin{tikzpicture}[
    block/.style={
        rectangle, draw, minimum width=2.5cm, minimum height=0.75cm,
        rounded corners, thick
    },
    scale=0.75,
    transform shape
]

\definecolor{ReLUColor}{RGB}{255, 165, 0}
\definecolor{WeightedLayerColor}{RGB}{173,216,230}
\definecolor{UpSampleColor}{RGB}{205,202,240}

\node[block, fill=WeightedLayerColor] (g2) at (2.5,0) {${W_{G}:F_{G} \times F}$};
\node[block, fill=WeightedLayerColor] (x2) at (2.5, -1.71) {${W_{x}:F_{x} \times F}$};

\node[circle, draw, thick, right=of g2, shift={(4, -0.85)}] (ew1) at (0, 0) {+};

\node[block, fill=ReLUColor, right=of ew1] (relu1) {ReLU};
\node[block, fill=WeightedLayerColor, right=of relu1] (wc1) {${W_{\psi}:F \times 1}$};
\node[block, right=of wc1, minimum width=1.25cm, minimum height=1.25cm] (sig1) {$\mathlarger{\mathlarger{\mathlarger{\mathlarger{\mathlarger{\sigma}}}}}$};
\node[block, right=of sig1, fill=UpSampleColor] (us3d1) {Upsample};
\node[circle, draw, thick, right=of us3d1] (ew2) {$\times$};

\foreach \i/\j in {ew1/relu1, relu1/wc1, wc1/sig1, sig1/us3d1, us3d1/ew2}
  \draw[-stealth, thick, shorten >=1pt] (\i) -- (\j);

\draw[-stealth, thick, shorten >=1pt] (0, 0) -- (g2);
\draw[-stealth, thick, shorten >=1pt] (0, -1.71) -- (x2);

\draw[-stealth, thick, shorten >=1pt]  (0.75, -1.71) -- node {} ++(0,-2cm) -| (ew2) node[pos=0.25] {} node[pos=0.75] {};

\draw[-stealth, thick, shorten >=1pt] (g2.east) -| (ew1.north);
\draw[-stealth, thick, shorten >=1pt] (x2.east) -| (ew1.south);

\coordinate (output) at (20.75, -0.85);

\draw[-stealth, thick, shorten >=1pt] (ew2.east) -- (output);

\node [left=0.35cm of g2, anchor=south] (t1) {G};
\node [left=0.35cm of x2, anchor=south] (t2) {x};

\node [above=0cm of sig1, anchor=south] {Sigmoid};

\coordinate (f1) at (6, -3.75);

% \begin{scope}
%     \node [draw, dashed, rounded corners, fit=(g2) (x2) (ew1) (wc1) (sig1) (us3d1) (ew2) (wc1) (sig1) (ew1) (in3) (output) (f1)] {};
% \end{scope}

\end{tikzpicture}
    \end{center}

    \caption{\textbf{Fig 5.} The Attention Gate (A) is from the Attention U-Net and computes the attention coefficient, $\alpha$, that maps the importance of the different spacial regions in the feature map.}
    
    \label{Fig:Model}
\end{figure*}

\subsection{Multiclass Segmentation Model}

The structure of the multiclass model can be seen in Figure 3. Due to the decreased in-memory size of the input after cropping the images, the multiclass segmentation model has an increased batch size of 6 instead of 2. The increased batch size allows for more aggressive data augmentations which increases the model's ability to generalize as the value of $\sigma$ for the elastic deformation data augmentation was uniformly distributed between 10 and 13. 

\quad The multiclass model had a similar structure to the binary model, however, it uses the ReLU activation function and employed 3 encoding and decoding blocks instead of 4 due to the smaller input dimensions compared to that of the binary model: dimensions of $128 \times 128 \times 128 \times 4$ were reduced to $48 \times 48 \times 128 \times 4$. Additionally, the number of filter channels was increased from 20 to 32 and soft attention mechanisms were implemented in the skip connections, similar to the attention U-Net, as well as in the encoding and decoding blocks, allowing the model to form more expressive connections within the data.

\quad In each encoding and decoding block, the model employs channel-based attention which determines the importance of each channel following the two convolutional layers as seen in Figure 4. The channel-based attention operator takes the global average of each channel which is then multiplied element-wise by a learnable parameter, $W_c \in \mathbb{R}^c$. The result is then fed into the sigmoid function, $\sigma$, to create the attention coefficients, $\alpha \in \mathbb{R}^c$ which consists of elements between 0 and 1 due to the sigmoid activation function. The channel-based attention block can be seen as follows:

\begin{equation}
\begin{aligned}
& \mu_{tc} = \frac{1}{HWD}\sum_{i}^{H}\sum_{j}^{W}\sum_{k}^{D}x_{tijkc}\\ 
& C(X) = \sigma (\mu_{ttc} \times W_c) \times X \\
\end{aligned}
\end{equation}

\noindent where $C(x)$ is the channel-based attention algorithm.

The model also employs attention between the skip connections through attention gates as in the Attention U-Net. The attention gate was based on human vision that focuses on a specific region and minimizes areas of little importance as in Figure 5. The attention gate algorithm, $A(X)$, can be seen where $G \in \mathbb{R}^{T \times W \times H \times D \times C}$ is the gating signal and $X \in \mathbb{R}^{T \times W \times H \times D \times C}$ is the input features (8). The linear transformations include $W_{X} \in \mathbb{R}^{F_{X} \times F}$, $W_{G} \in \mathbb{R}^{F_{G} \times F}$, $W_{\Psi} \in \mathbb{R}^{F \times 1}$ and the attention gate algorithm is as follows: 

\begin{equation}
\begin{aligned}
& \alpha = \sigma (W_{\psi}\times ReLU(W_{X} \times X + W_{G} \times G) \\
& A(X) = \alpha \times X \\
\end{aligned}
\end{equation}

\noindent where $\alpha$ represents the attention coefficient that produces the relevant and irrelevant feature map when element-wise multiplied with $X$, the input features \cite{b8}. The overall architecture of the multiclass segmentation model is fairly basic as the region of interest detection algorithm sufficiently decreased the complexity of the task by reducing the high-dimensional input.

\subsection{Training}
Multiple loss functions were implemented and tested during the training of the multiclass segmentation model. These loss functions include categorical cross-entropy ($\mathcal{L}_{CE}$), dice loss ($\mathcal{L}_{DL}$), Log-Cosh Dice Loss ($\mathcal{L}_{LCDL}$), and a combination of dice loss and cross-entropy ($\mathcal{L}_{DLCE}$) as follows \cite{b17}:

\begin{equation}
\mathcal{L}_{CE} = CE(y,\hat{y}) = \begin{cases} \\
-log(\hat{p}), & y=1 \\
-log(1-\hat{p}), & \text{otherwise} \\
\end{cases} \\
\end{equation}

\begin{equation}
\mathcal{L}_{LCDL} = CE(y,\hat{y}) = log(\frac{e^{\mathcal{L}_{DL}}+e^{-{\mathcal{L}_{DL}}}}{2})
\end{equation}

\begin{equation}
\mathcal{L}_{DLCE} = \mathcal{L}_{DL}+\mathcal{L}_{CE}
\end{equation}

Furthermore, the model was trained using different optimizers including Adam, RAdam, and Lookahead, which can be seen in Algorithm 1 \cite{b20}. The key difference between Adam and RAdam is the presence of an additional term that attempts to correct the large variability of the adaptive learning rate in the early stages of learning \cite{b21}.

\begin{algorithm}
\caption{Lookahead Optimizer}
\begin{algorithmic}[1]
    \REQUIRE {Parameters $\phi$, loss function $\mathcal{L}$}
    \REQUIRE {Synchronization period K, slow weights step size $\alpha$, optimizer A}
    \FOR {$t=1,2,\ldots$}
    \STATE {Synchronize parameters $\theta_{t,0}\gets \phi_{t-1}$}
    \FOR {$i=1,2,\ldots, k$}
    \STATE {sample minibatch of data $d\sim D$}
    \STATE $\theta_{t,i}\gets \theta_{t,i-1} + A(\mathcal{L},\theta_{t,i-1},d)$
    \ENDFOR
    \STATE $\phi_{t}\gets \phi_{t-1} + \alpha(\theta_{t,k}-\phi_{t-1})$
    \ENDFOR
\end{algorithmic}
\end{algorithm}

The model was trained with a learning rate of 0.0003 using the different previously mentioned loss functions and optimizers. The model was trained for 300 epochs with a batch size of 6. The architecture of the multiclass segmentation model can be seen in Figure 3. 

\subsection{Test-Time Augmentations}

Data augmentations were initially employed to increase the size of the training dataset. Common data transformations include reflection, rotation, shear, and scaling images. Test-time augmentations apply these augmentations on the test set instead of the training data to create different predictions on the single image. These different predictions of the transformed images are then aggregated. Prior studies show that combining multiple predictions was previously used to increase performance for biomedical tasks \cite{b5}. For instance, Matsunaga et al and Radosavocic implemented test-time augmentations for skin lesion classification and pulmonary node detection, respectively \cite{b22,b23}.
\subsection{Energy-Based Model}
Instead of using the softmax confidence score of the different predictions, an energy-based model was implemented. The objective of the energy-based model is to create a function $E(x): \mathbb{R^{K}}\to \mathbb{R}$ that maps the logit values from the probability density function to a non-probabilistic scalar known as energy. Energy-based models are more reliable than the softmax confidence score due to the increased disparity between in- and out-distribution samples. The energy-based model is as follows: 
\begin{equation}
E(x;f) = -log\sum_{i}^{K}e^{f_{i}(x)}
\end{equation}
\noindent shows K as the number of classes, $f(x) : \mathbb{R}^{D}\to \mathbb{R}^{K}$ as the neural network, and $x\in \mathbb{R}^{D}$ as the different 3D MRI modalities.
Liu et al showed the mathematical connection between the energy model and the softmax confidence as follows \cite{b24}: 
\begin{equation}
\begin{aligned}
& \max_{y}p(y|x)=\max_{y}\frac{e^{f_{y}(x)}}{\sum_{i}^{K}e^{f_{i}(x)}}=\frac{e^{f^{max}(x)}}{\sum_{i}^{K}e^{f_{i}(x)}} \\
& =\frac{1}{\sum_{i}^{K}e^{f_{i}(x)-f^{max}(x)}} \\
& log(\max_{y}p(y|x))=E(x;f(x)-f^{max}(x)) \\
& =\underbrace{E(x;f)} _{\downarrow\mathrm{\text{ for in-dist x}}}+\underbrace{f^{max}(x)} _{\uparrow\mathrm{\text{ for in-dist x}}} \\
\end{aligned}
\end{equation}
Using the energy-based model, each voxel of the predicted mask was replaced with an energy score. The final uncertainty prediction was $-E(x;f)$ to follow the convention that high energy scores represent certain predictions, while low energy scores show uncertain predictions. 

\section{Results}
\subsection{Data Description}\label{AA}

In this study, we used the BraTS 2020 dataset to train the proposed models. The BraTS 2020 dataset contains multiple 3D MRI modalities with various annotated histological regions each with different sizes and aggressiveness \cite{b25,b26}. Each MRI modality has dimensions of $240 \times 240 \times 155$ with a 1mm isotropic resolution. These MRIs were clinically obtained using 3T multimodal MRI scans for T1, T1-Gd, T2, and FLAIR MRI scans.

\quad The dataset is comprised of 369 gliomas, including 259 cases of high-grade gliomas (HGG) and 110 cases of low-grade gliomas (LGG). Four MRI modalities, T1, T1-Gd, T2, and FLAIR were collected from each patient \cite{b28,b27,b28}. Neuro-radiologists then labeled the histological regions (peritumoral edema, enhancing tumor, and necrotic non-enhancing tumor core) along with normal brain tissue. The dataset was then split into training (80\%) and test (20\%) sets. The experimental results were achieved using an Intel Core I7 processor, 64 GB RAM, and an NVIDIA 3090 GPU. The models were tested using the Dice similarity metric which computes the overlap between the ground truth and the prediction. The metric is as follows:
\begin{equation}
\text{Dice}(P,T)=2\times \frac{P\cap T}{P+T} \\
\end{equation}
shows where P and T represent the predicted masks and the ground truth, respectively \cite{b9}.
\subsection{Segmentation Results (Quantitative)}
Various methods and model architectures were proposed including strided convolutions instead of max pooling (SC), instance normalizations (I-Norm), and the region of interest detection algorithm (ROI) which was added onto a U-Net with 3 encoding and decoding blocks. The loss functions that were implemented in the experiment include: dice loss (DL), log cosh dice loss (LC), categorical cross-entropy (CE), and a combination of dice loss and categorical cross-entropy (CE+DL). Lastly, the implemented optimizers include RAdam (RA), Adam (A), Ranger (R), and Adam and Lookahead (A+LH). The performance of the different loss functions and optimizers can be seen in Table I with the different proposed methods (SC, IN, ROI). The order of the ablation study tested the different loss functions on the Adam optimizer. The optimizers were tested using the highest-performing loss function, log cosh dice loss. Using the highest-performing optimizer and loss function, Table II shows the results of the different proposed methods. Each of the methods, instance normalization, strided convolutions, and the region of interest detection algorithm, increased the model's performance as seen in a higher mean dice score.

\quad To further analyze the model’s performance, we compared its results to other popular segmentation models that were trained and tested on the BraTS2020 dataset. These models include the U-Net, Isensee et al’s nnUnet, Fidon et al’s nnUnet ensemble, Jia et al’s H2NF-Net, Raza et al's dResU-Net, and Liu et al's ADHDC-Net \cite{b29,b30,b31, b32, b33}. The quantitative results can be seen in Table III. From Table III, the proposed network and region of interest detection algorithm achieved the greatest mean dice score. Specifically, the model outperformed its contemporaries when segmenting the enhancing and non-enhancing tumor regions. 

\quad Since the proposed models and the nnUnets had a relatively similar architecture, it was inferred that the enhanced and non-enhancing regions' improved performance was due to the region of interest detection algorithm, along with the added soft attention mechanisms.

\begin{table}
\begin{center}
\caption{Table I: Compared Loss and Optimizers}
\begin{tabular}{ l c c c c}
\toprule
Segmentation Method & Whole & Core & Enh. & Mean \\
\midrule
DL+A & 89.62 & 89.2 & 85.54 & 88.11 \\
CE+A & 86.22 & 84.9 & 79.11 & 83.41 \\
DL+CE+A & 89.62 & \textbf{89.8} & 84.89 & 88.10 \\
LC+A & \textbf{90.13} & 89.14 & 85.85 & 88.37 \\
LC+R & 89.9 & 89.37 & 85.89 & 88.39 \\
LC+RA & 90.12 & 89.45 & 85.62 & 88.4 \\
LC+A+LH & 89.13 & 88.97 & \textbf{87.47} & \textbf{88.52} \\
\bottomrule
\end{tabular}
\end{center}
\end{table}

\begin{table}
\begin{center}
\caption{Table II: Compared Methods}
\begin{tabular}{ l c c c c}
\toprule
Segmentation Method & Whole & Core & Enh. & Mean \\
\midrule
U-Net & 84.72 & 86.4 & 77.42 & 82.85 \\
ROI & 88.65 & 89.11 & 82 & 86.83 \\
ROI+SC & 88.92 & \textbf{89.15} & 84.53 & 87.53 \\
ROI+SC+I-Norm & \textbf{89.13} & 88.97 & \textbf{87.47} & \textbf{88.52} \\
\bottomrule
\end{tabular}
\end{center}
\end{table}

\begin{table}
    \centering
    \caption{Table III: Compared to SOTA on BraTS 2020}
    \begin{tabular}{l c c c c}
        \toprule
        Method & Whole & Core & Enh. & Mean \\
        \midrule
        nnUnet \cite{b29} & 91.07 & 87.97 & 81.37 & 87.07 \\
        H2NF-Net \cite{b30} & \textbf{91.3} & 85.5 & 78.8 & 86.2 \\
        nnUnet Ensemble \cite{b31} & 91.0 & 84.4 & 77.6 & 84.8 \\
        dResU-Net \cite{b32} & 86.60 & 83.57 & 80.04 & 83.40 \\
        ADHDC-Net \cite{b33} & 89.75 & 83.31 & 78.01 & 83.69 \\
        Proposed & 89.13 & \textbf{88.97} & \textbf{87.47} & \textbf{88.52} \\
        \bottomrule
    \end{tabular}
\end{table}

\subsection{Segmentation Results (Qualitative)}
The images of the test set were ranked based on the mean dice score of the 3 regions and the 0\textsuperscript{th}, 25\textsuperscript{th}, 50\textsuperscript{th}, 75\textsuperscript{th}, and 100\textsuperscript{th} percentile predictions can be seen in Figure 6. Based on the shown figure, the overall performance of the model is high.  For non-performant predictions, the mean dice score is low due to the poor performance in segmenting the enhancing and non-enhancing tumor regions; due to the small size of these regions, small errors can lead to very low dice scores. Although the performance in these regions is low, the uncertainty map shows low confidence due to the severely low energy score. 

\quad For the 25\textsuperscript{th}, 50\textsuperscript{th}, and 75\textsuperscript{th} percentile predictions, the predictions are far more accurate, however, they still have a few areas of inaccuracy. In these regions of error, the uncertainty map shows low confidence. On the other hand, in the accurate areas, the uncertainty map shows high confidence. Lastly, in the 100\textsuperscript{th} percentile prediction, the model's performance is very accurate with the uncertainty map indicating high confidence. Overall, the uncertainty methods are accurate because they show high confidence when accurate and low confidence when inaccurate.

\subsection{Additional Testing}
To further test the robustness of the proposed model, we measured its performance on other datasets: BraTS 2019 and BraTS 2021. These datasets are similar to BraTS 2020 as they all contain the 3D MRI modalities (T1, T1Gd, T2, and FLAIR) and annotated histological regions. The BraTS 2019 dataset contains 335 gliomas from patients including 259 HGGs and 76 LGGs and the BraTS 2021 dataset contains 1251 gliomas. These gliomas are captured using 3D MRIs that were skull stripped and resampled to $1mm^3$  \cite{b25, b26, b27, b28, b34, b35}. The BraTS 2019 and 2021 datasets were split using the same process as the BraTS 2020 dataset. 

\quad Similar to the testing of the BraTS 2020 dataset, we also compared our model's performance on the BraTS 2019 and 2021 datasets to other state-of-the-art (SOTA) models. These models include Jiang et al’s cascaded U-Net, Zhao et al’s naive U-Net, McKinley et al’s DeepSCAN, Rehman et al's RAAGR2-Net, Futrega et al's DynUnet, Pham et als SegTransVAE, Hatamizadeh et al's Swin UNETR, Liang et al's 3D PSwinBTS, Liu et al's ADHDC-Net, and Lin et al's CKD-TransBTS \cite{b33,b36,b37,b38,b39,b40,b41,b42,b43,b44}. The SOTA models vary from the different BraTS datasets as we specifically picked the best-performing models for each dataset, rather than high-performing models tested on all three BraTS datasets. The performance of the proposed model compared to its contemporaries can be seen in Table IV and V for the BraTS 2019 and 2021 datasets, respectively. For the BraTS 2019 dataset, the proposed model outperformed other state-of-the-art models when segmenting the core and enhancing regions, while also having the second-highest mean dice score. Additionally, the proposed model outperformed its contemporaries on the BraTS 2021 dataset. 

\quad The increased performance of the proposed model was due to the implemented methods: strided convolution, instance normalization, region of interest detection algorithm, attention gate, and channel-wise attention. The strided convolutions replaced max pooling which decreased data loss, allowing the model to obtain a better representation of the brain MRIs. Additionally, the instance normalization allowed the model to converge faster and be more robust as seen in the ablation studies. The attention mechanisms were believed to have the largest effect on the models performance. The attention gate caused the skip connections to primary focus on the important features. The channel-wise attention convolutional block implemented throughout the U-Net allowed for better representations within the channels in the encoder and decoder. Although the model had a consistently worse performance on the peritumoral task, it performed substantially better on the enhancing and non-enhancing regions. This leveled performance among classes was believed to be due to the region of interest as it counteracted the class imbalance.

\begin{table}
    \centering
    \caption{Table IV: Compared to SOTA on BraTS 2019}
    \begin{tabular}{l c c c c}
        \toprule
        Method & Whole & Core & Enh. & Mean \\
        \midrule
        ADHDC-Net \cite{b33} & 89.94 & 83.89 & 77.91 & 83.91 \\
        Cascaded U-Net \cite{b36} & 90.9 & 86.5 & 80.2 & \textbf{85.87} \\
        Naive U-Net \cite{b37} & \textbf{91.0} & 83.5 & 75.4 & 83.3 \\
        DeepSCAN \cite{b38} & \textbf{91} & 83 & 77 & 83.67 \\
        RAAGR2-Net \cite{b39} & 89.6 & 82.1 & 77.6 & 82.2 \\
        Proposed & 84.39 & \textbf{86.51} & \textbf{82.74} & 84.55 \\
        \bottomrule
    \end{tabular}
\end{table}

\begin{table}
    \centering
    \caption{Table V: Compared to SOTA on BraTS 2021}
    \begin{tabular}{l c c c c}
        \toprule
        Method & Whole & Core & Enh. & Mean \\
        \midrule
        DynUnet \cite{b40} & 92.88 & 89.71 & 85.81 & 89.46 \\
        SegTransVAE \cite{b41} & 92.54 & 89.99 & 86.22 & 89.58 \\
        Swin UNETR \cite{b42} & 92.73 & 89.98 & 86.81 & 89.84 \\
        PSwinBTS \cite{b43} & \textbf{93.62} & 90.43 & 88.25 & 90.76 \\
        CKD-TransBTS \cite{b44} & 93.33 & 90.16 & 88.50 & 90.66 \\
        Proposed & 90.56 & \textbf{92.12} & \textbf{89.77} & \textbf{90.82} \\
        \bottomrule
    \end{tabular}
\end{table}

\begin{figure*}[htbp]
    \centering
    \begin{tabular}{cccccc}
        &
        \small 0th Percentile &
        \small 25th Percentile &
        \small 50th Percentile &
        \small 75th Percentile &
        \small 100th Percentile \\

        \rotatebox{90}{\small \quad \quad \quad FLAIR} &
        \includegraphics[width=0.17\linewidth]{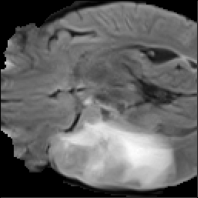} & 
        \includegraphics[width=0.17\linewidth]{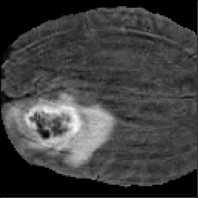} & 
        \includegraphics[width=0.17\linewidth]{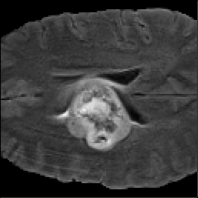} & 
        \includegraphics[width=0.17\linewidth]{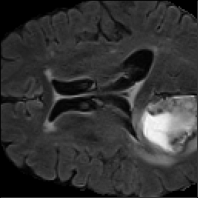} & 
        \includegraphics[width=0.17\linewidth]{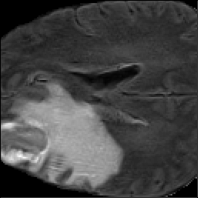} \\

        \rotatebox{90}{\small \quad \quad \quad \space T1-Gd} &
        \includegraphics[width=0.17\linewidth]{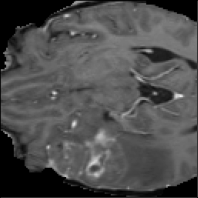} & 
        \includegraphics[width=0.17\linewidth]{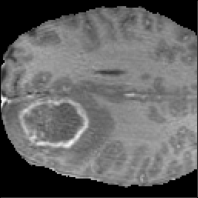} & 
        \includegraphics[width=0.17\linewidth]{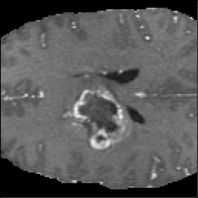} & 
        \includegraphics[width=0.17\linewidth]{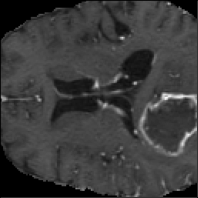} & 
        \includegraphics[width=0.17\linewidth]{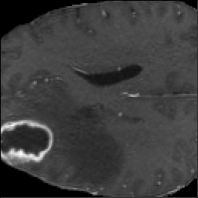} \\

        \rotatebox{90}{\small \quad \quad \space Prediction} &
        \includegraphics[width=0.17\linewidth]{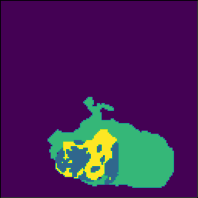} & 
        \includegraphics[width=0.17\linewidth]{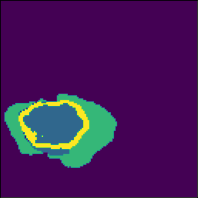} & 
        \includegraphics[width=0.17\linewidth]{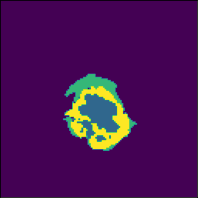} & 
        \includegraphics[width=0.17\linewidth]{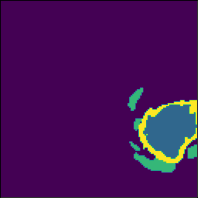} & 
        \includegraphics[width=0.17\linewidth]{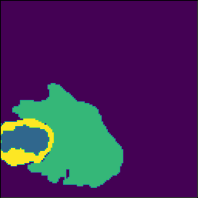} \\

        \rotatebox{90}{\small \quad \space \space Ground Truth} &
        \includegraphics[width=0.17\linewidth]{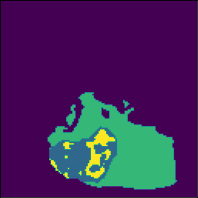} & 
        \includegraphics[width=0.17\linewidth]{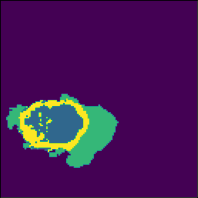} & 
        \includegraphics[width=0.17\linewidth]{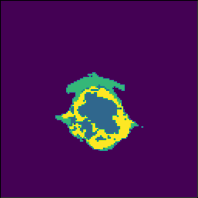} & 
        \includegraphics[width=0.17\linewidth]{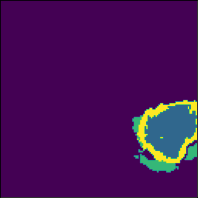} & 
        \includegraphics[width=0.17\linewidth]{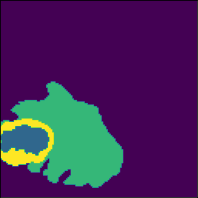} \\

        \rotatebox{90}{\small \quad \space \quad Uncertainty} &
        \includegraphics[width=0.17\linewidth]{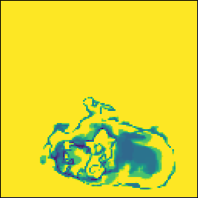} & 
        \includegraphics[width=0.17\linewidth]{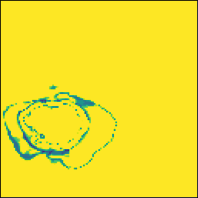} & 
        \includegraphics[width=0.17\linewidth]{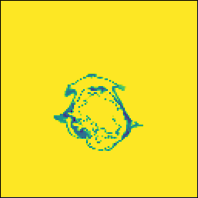} & 
        \includegraphics[width=0.17\linewidth]{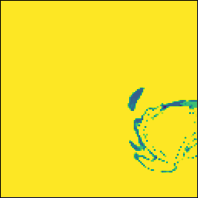} & 
        \includegraphics[width=0.17\linewidth]{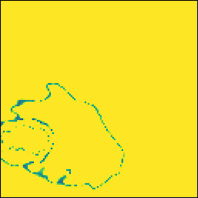} \\
    \end{tabular}
    
    \vspace{4pt}
    
    \caption {\textbf{Fig 6.} Qualitative Brain Tumor Segmentation results on BraTS 2020. Shown are the 0\textsuperscript{th}, 25\textsuperscript{th}, 50\textsuperscript{th}, 75\textsuperscript{th}, and 100\textsuperscript{th} predictions. For the prediction and ground truth, the green, yellow, and blue regions indicate the peritumoral edema, enhancing tumor, and non-enhancing tumor region, respectively. On the bottom is the uncertainty map where yellow and non-yellow regions show confidence and confidence, respectively.}
    \label{Fig:Model}
\end{figure*}

\section{Discussions}

This paper proposes a novel deep-learning framework for brain tumor segmentation, as well as a region of interest detection algorithm. These frameworks employ fully convolutional autoencoders as well as soft and hard attention mechanisms. Experimental results demonstrated the efficacy of the proposed methods, specifically, in their ability to treat class imbalances. Furthermore, the proposed framework achieves better segmentation results when compared to other state-of-the-art methods that have been tested on the BraTS benchmarks including BraTS 2019, 2020, and 2021. Future works would explore the model's feasibility in a clinical setting. Although the framework performed well on the datasets provided, additional patient MRIs from different facilities and conditions would be needed to validate the framework’s strong performance and robustness in practice. Additionally, we would explore extending the proposed model to other semantic segmentation problems in the medical field. Regarding the uncertainty estimations, the employed test-time augmentations and energy-based model showed strong results supported by the model’s high confidence when accurate and low confidence when inaccurate. One limitation of the uncertainty methods was the lack of quantitative data to show its efficacy. As a result, its performance was only evaluated qualitatively. Future improvements to the uncertainty estimation methods would include incorporating other transformations other than reflection. Based on the model’s performance across the different BraTS benchmarks, we strongly believe that it can be greatly improved through more training data. The segmentation results on BraTS 2021 were substantially better than those of BraTS 2019 and 2020. However, we also believe that the model's performance is bottlenecked by the high variability in brain scans along with potential subjectivity in the annotations. 

\section{Conclusions}

In this paper, we proposed various algorithms and models for the region of interest detection and multiclass segmentation of high and low-grade gliomas. The ROI cropping heavily decreased computation and was believed to aid in reducing the class imbalance by increasing the performance of the enhancing and non-enhancing tumor regions. Furthermore, the proposed model for multiclass segmentation achieved high results as it outperformed other state-of-the-art models on BraTS benchmarks. Lastly, the voxel-based uncertainty estimation techniques provided a visual confidence map. This allows for effective clinical implementation by allowing doctors to ignore unconfident and inaccurate predictions while trusting confident and accurate ones. In a clinical implementation, it is believed that the segmentation pipeline and uncertainty estimation would heavily decrease the current error rate of diagnosing brain tumors, 20\%, while creating an accurate and automated method of segmentation. \cite{b3}.

\bibliographystyle{model1-num-names.bst}
\bibliography{main}

\end{document}